\documentclass[%
 showpacs,
 reprint,
 amssymb,amsmath,%
aip,
jap,
]{revtex4-1}

\usepackage[latin1]{inputenc}

\newcommand{\diff}[0]{\text{d}}
\newcommand{\re}[0]{\text{\,Re}}

\usepackage{graphicx}
\usepackage{dcolumn}
\usepackage{bm}
\usepackage{color}

\begin{document}

\preprint{APS/123-QED}

\title{Model for reflection and transmission matrices of nanowire end facets}

\author{Guro K. Svendsen}
\author{Helge Weman}%
\author{Johannes Skaar}%
\altaffiliation[Also at ]{University Graduate Center, Kjeller, Norway.}

\affiliation{Department of Electronics and Telecommunications, Norwegian University of Science and Technology, NO-7491 Trondheim, Norway}
%


\date{\today}

\begin{abstract}
Nanowires show a large potential for various electrooptical devices, such as light emitting diodes, solar cells and nanowire lasers. We present a method developed to calculate the modal reflection and transmission matrix at the end facets of a waveguide of arbitrary cross-section, resulting in a generalized version of the Fresnel equations. The reflection can be conveniently computed using \textit{Fast Fourier Transforms}, once the waveguide modes are known. We demonstrate that the reflection coefficient is qualitatively described by two main parameters, the modal field confinement and the average Fresnel reflection of the plane waves constituting the waveguide mode.  
\end{abstract}


\maketitle

\section{Introduction}
Nanowires have received significant attention due to promising applications in electrooptical devices. Nanowire lasers \cite{Duan2003241}, light emitting diodes \cite{lieber:led} and solar cells \cite{lieber} have already been demonstrated. Semiconductor nanowires are characterized by large aspect ratios and high index contrast. Various techniques can be used to accurately control the composition and geometry of the nanowires. Some work has been done to analyze the optical properties of semiconductor nanowires \cite{maslov,henneghien,maslov:farField,maslov:optEmission,chen:coupOptEl,maslov:modalGain,hauschild}, including  modal properties, modal gain and optical emission. More knowledge is however needed to better understand and predict the optical response of the nanowires, in order to improve the nanowire devices. For nanowire lasers, two especially important properties are the reflection and transmission at the end facets. The reflection is important for the laser resonator, whereas the transmission yields the far field radiation.

The reflection properties of nanowires have been investigated by \textit{Maslov et al.} \cite{maslov} and \textit{Henneghien et al.} \cite{henneghien}, using a 3d finite element method, and also by \textit{Hugonin et al.}
\cite{hugonin} using coupled-wave analysis \cite{friedler}. With this work they have contributed with important knowledge concerning the reflection properties of nanowires. Calculation of the reflection and transmission at an interface between a waveguide and some half-space is however a 2d problem connecting the true waveguide modes at each side of the interface; thus it is not necessary to perform a full 3d analysis. In addition, the cross coupling between different modes, including coupling to radiation modes, and the transmission far field, are not obtained directly with the finite element methods.

In planar dielectric waveguides, various techniques have been used to analyze step discontinuities \cite{marcuse,suchoski,vassallo, kendall,wu,amditis,vahidpour}. Several of the techniques are based on matching modal fields at the boundary \cite{vassallo, kendall, wu, amditis, marcuse, suchoski}. Most attention has been given to conventional waveguide structures with low index contrast.

In this work we use the electromagnetic boundary conditions to obtain matrix Fresnel equations. These equations directly give the modal reflection matrix $\left[r_{ij}\right]$, and the transmission matrix $[t_{im}]$. The model has mainly been intended for calculations on nanowires, but is also useful for other waveguide geometries. A key aspect of nanowires are their small lateral size, often on the scale of the light wavelength, leading to significant diffraction effects. In our model, these diffraction effects are manifested as cross coupling to other modes, including radiation modes. The model is more compact and simple in its form than the previous 1d techniques, and it is directly applicable to waveguides of arbitrary cross-section, both in 1d (planar waveguides) and 2d.

This article is organized as follows. The general reflection model for an arbitrary geometry is discussed in Section II. In Section III the special case of a planar waveguide is presented. The simplifications involved in the planar case facilitate the illustration of some important aspects of the general model. Results for hexagonal nanowires are presented in Section IV: a wide bandgap nanowire based on a material similar to ZnO or GaN, a narrow bandgap pure GaAs nanowire and heterostructured GaAs nanowires with Al$_{0.3}$Ga$_{0.7}$As claddings. Finally, in Section V, the model is used to calculate the reflection properties of a multimode optical fiber.

\section{Calculating the reflection and transmission matrices}
\label{sec:calculating_r_t}
\begin{figure}[bt]
	\centering
	\includegraphics{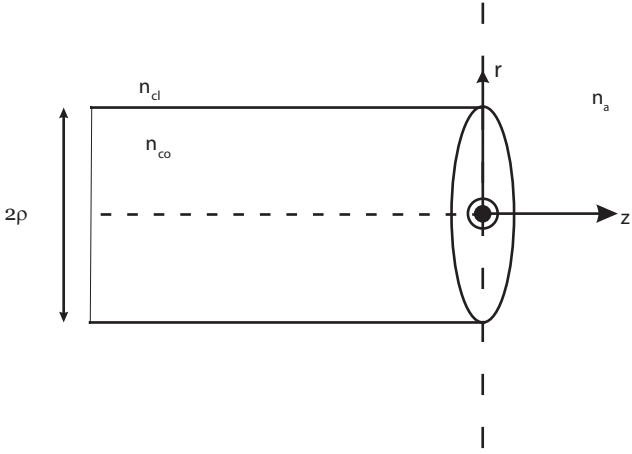}
	\caption{Waveguide oriented along the $z$-axis with end facet at ${z=0}$. The waveguide core and cladding have refractive indices $n_{\text{co}}$ and $n_{\text{cl}}$, respectively. The ambient medium for $z>0$ is assumed to be homogeneous with refractive index $n_{\text{a}}$. The fields are discretized using periodic boundary conditions, i.e. the structure is assumed to be periodic with periodicity $2L$ in both $x$ and $y$ direction, where $2L\gg2\rho$ .}
	\label{fig:circ_wg}
\end{figure}

We will now demonstrate how the electromagnetic boundary conditions may be reformulated as matrix Fresnel equations. The waveguide is oriented along the $z$-axis and terminated at $z=0$, see Fig. \ref{fig:circ_wg}. No particular geometry is assumed for the cross-section. At each side of the boundary, $z=0$, the mode spectrum is discretized. The discretization can be done using e.g. periodic boundary conditions, i.e. the single waveguide is approximated by a periodic array of nanowires. We surround the waveguide with a quadratic box of size $(2L)^2$, and let the area within this box constitute the unit cell of the periodic grating. The period in both $x$- and $y$-direction is thus $2L$. The parameter $L$ should be large enough, so that neighboring waveguides do not perturb the field of the bound waveguide modes. In the ambient half-space $z>0$ there is a homogeneous isotropic medium with refractive index $n_{\text{a}}$. The electric field of the modes can be written
\begin{align}
&\mathbf{\mathcal{E}}_{m}(x,y)= \mathbf{\hat{\mathcal{E}}}_{m}\exp(ik_xx+ik_yy).\label{modeexpE}
\end{align}
The magnetic field, $\mathbf{\mathcal{H}}_{m}$, is described in the same way. Each mode is labeled by a collective mode index $m=m(k_x,k_y,\textrm{pol})$, specifying the wavevector and the polarization. The polarization, pol, is either TE or TM. The real transverse wavevectors are $k_x=p\frac{\pi}{L}$ and $k_y=q\frac{\pi}{L}$, where $p$ and $q$ are positive integers. The modal propagation constant $k_z$ is given by $k_z^2=n_{\text{a}}^2\omega^2/c^2-k_x^2-k_y^2$, where $n_{\text{a}}$ is the refractive index of the half-space $z>0$, and $c$ is the vacuum light velocity.  
The four constant vectors can be expressed \cite{Note1}

\begin{subequations}\label{constvectors}
\begin{align}
&\mathbf{\hat {\mathcal{E}}}_{m(\text{TE})}=A\sqrt{\omega\mu}(-k_y,k_x,0),\label{eTE}\\
&\mathbf{\hat {\mathcal{E}}}_{m(\text{TM})}=\frac{A}{\sqrt{\omega\varepsilon_{\text{a}}}}\left(k_xk_z,k_yk_z,-\left(k_x^2+k_y^2\right)\right),\label{eTM}\\
&\mathbf{\hat {\mathcal{H}}}_{m(\text{TE})}=\frac{A}{\sqrt{\omega\mu}}(-k_xk_z,-k_yk_z,k_x^2+k_y^2),\label{hTE} \\
&\mathbf{\hat {\mathcal{H}}}_{m(\text{TM})}=A\sqrt{\omega\varepsilon_{\text{a}}}(-k_y,k_x,0)\label{hTM},
\end{align}
\end{subequations}
where $A=1\left/{\sqrt{(k_x^2+k_y^2)|k_z|2L^2}}\right.$, $\varepsilon_{\text{a}}$ is the permittivity, and $\mu$ is the  permeability. From Maxwell's curl equations, \eqref{hTE} and \eqref{eTM} follow from \eqref{eTE} and \eqref{hTM}, respectively. The modes are orthogonal in the sense
\begin{align}
&\frac{1}{2}\int\limits_{\text{cell}} \mathbf{\mathcal{E}}_{m}(x,y)\times\mathbf{\mathcal{H}}_{m'}^*(x,y)\cdot\mathbf{\hat z} \mathrm{d}A=\frac{\kappa(m)}{|\kappa(m)|}\delta_{m,m'}
\label{orthogonal}.
\end{align}
Here, the area of integration is defined by $-L\leq{x}\leq{L}$ and $-L\leq{y}\leq{L}$.
The unit vector in the $z$-direction is $\mathbf{\hat z}$, and
\begin{equation}\kappa(m)=\left\{ \begin{smallmatrix} 
k_z^*,\:\text{pol}=\text{TE}\\ 
k_z,\:\text{pol}=\text{TM}. \end{smallmatrix} \right.\end{equation}

For the waveguide ($z<0$), the modes can be divided into two sets: Discrete bound modes, and continuous radiation modes \cite{snyder}. Due to periodic boundary conditions (or alternatively, metallic boundary conditions), however, the continuous set of radiation modes is converted into a discrete set. We will accordingly describe the modal fields by the discrete set $\{\mathbf e_i,\mathbf h_i\}$, where ${\mathbf e_i=\mathbf{e}_i(x,y)}$ and $\mathbf h_i=\mathbf h_i(x,y)$ are the electric and magnetic fields, respectively, of mode $i$.

The transversal electric field must be continuous at the boundary. Assuming incoming mode $\mathbf e_i(x,y)$ we can write
\begin{align}
&\mathbf e_i^{(t)}+\sum_j r_{ij}\mathbf e_j^{(t)}=\sum_{m} t_{im}\mathbf{\mathcal{E}}_{m}^{(t)}, \label{contE}
\end{align}
valid for all $x$ and $y$. Here $r_{ij}$ is the reflection coefficient from mode $i$ to mode $j$, and $t_{im}$ is the transmission coefficient from mode $i$ (in the waveguide) to mode $m$ (in the half-space $z>0$). The superscript $(t)$ stands for the transversal component ($x$ and $y$ components) of the vector. The boundary condition for the transversal magnetic field is similarly
\begin{align}
&\mathbf h_i^{(t)}-\sum_j r_{ij}\mathbf h_j^{(t)}=\sum_{m} t_{im}\mathbf{\mathcal{H}}_{m}^{(t)}\label{contH}.
\end{align}
Eqs.~\eqref{contE} and \eqref{contH} can be combined as follows. Take the vector product between \eqref{contE} and $\mathbf{\mathcal{H}}_{m'}^{(t)*}(x,y)$, and integrate over the unit cell. Similarly, take the vector product between $\mathbf{\mathcal{E}}_{m'}^{(t)*}(x,y)$ and \eqref{contH}, and integrate over the unit cell. This yields
\begin{subequations}\label{transmission}
\begin{align}
&t_{im}=\Psi_{im}+\sum_j r_{ij}\Psi_{jm}\label{transmission_e}\\
&t_{im}=\Phi_{im}-\sum_j r_{ij}\Phi_{jm},\label{transmission_h} 
\end{align}
\end{subequations}
where
\begin{subequations}\label{innerprod}
\begin{align}
&\Psi_{im}=\frac{\kappa(m)^*}{|\kappa(m)|}\frac{1}{2}\int\limits_{\text{cell}}\mathbf e_i^{(t)}\times\mathbf{\mathcal{H}}_{m}^{(t)*}(x,y)\cdot\mathbf{\hat z}\mathrm{d}A, \\
&\Phi_{im}^*=\frac{\kappa(m)^*}{|\kappa(m)|}\frac{1}{2}\int\limits_{\text{cell}}\mathbf{\mathcal{E}}_{m}^{(t)}(x,y)\times\mathbf h_i^{(t)*}\cdot\mathbf{\hat z}\mathrm{d}A.
\end{align}
\end{subequations}
Eliminating $t_{im}$, we obtain
\begin{align}
&\Psi_{im}+\sum_j r_{ij}\Psi_{jm} =\Phi_{im}-\sum_j r_{ij}\Phi_{jm}.\label{matreq}
\end{align}
Eqs.~\eqref{transmission} and \eqref{matreq} are multimode, matrix versions of the electromagnetic boundary conditions. The matrices $[\Psi_{im}]$ and $[\Phi_{im}]$ have first index $i$ (designating the waveguide mode) and second index $m$ (designating the mode of the homogeneous medium). Solving \eqref{matreq} yields the matrix form of Fresnel's equation to calculate the reflection matrix:
\begin{equation}\label{fresnel}
[r_{ij}]=([\Phi_{im}]-[\Psi_{im}])([\Phi_{im}]+[\Psi_{im}])^{+}.
\end{equation}
In \eqref{fresnel}, we have assumed that there are at least as many $m$ values as the number of $i$'s; thus the equation can be solved with the help of Moore--Penrose pseudoinverse (denoted by superscript $+$). Once the reflection matrix is found, the transmission matrix can be calculated with the help of \eqref{transmission}.

The matrices $\Phi_{im}$ and $\Psi_{im}$ can be written out in terms of the Cartesian components of the modal fields as follows:
\begin{subequations}\label{innerprod_psi_phi}
\begin{align}
&\Psi_{im(\text{TE})}=\frac{A|k_z|}{2\sqrt{\omega \mu}}\left(-{\widetilde{e_i}^{x}}k_y+{\widetilde{e_i}^{y}}k_x\right),\\
&\Psi_{im(\text{TM})}=\frac{A|k_z|\sqrt{\omega\varepsilon_{\text{a}}}}{2k_z}\left({\widetilde{e_i}^{x}}k_x+{\widetilde{e_i}^{y}}{k_y}\right),\\
&\Phi_{im(\text{TE})}=\frac{-A|k_z|\sqrt{\omega \mu}}{2k_z}\left({\widetilde{h_i}^{y}}{k_y}+{\widetilde{h_i}^{x}}{k_x}\right),\\
&\Phi_{im(\text{TM})}=\frac{A|k_z| }{2\sqrt{\omega\varepsilon_{\text{a}}}}\left({\widetilde{h_i}^{y}}{k_x}-{\widetilde{h_i}^{x}}{k_y}\right).
\end{align}
\end{subequations}
Here we have defined
\begin{align}
\label{fourier}
{\widetilde{e_j}^{(x)}}=\int\limits_{\text{cell}} e_{j}^{(x)}\exp(-ik_x x- ik_y y)\mathrm{d}A.
\end{align}
The other field components are defined similarly. 

The distance between neighboring transverse wavevectors must be sufficiently small, so that a superposition of the associated modes in the half-space $z>0$ can describe the modes of the waveguide ($z<0$) accurately. In other words, we require the spacing
\begin{equation}
\Delta k_x < \frac{\pi}{\mathcal{L}},\ \Delta k_y < \frac{\pi}{\mathcal{L}},\label{delta_k}
\end{equation}
where $2\mathcal{L}$ is the maximum extension of the bound modes. Ineq.~\eqref{delta_k} will be fulfilled for $L>\mathcal{L}$. 

For practical computation, we first need to compute the waveguide modes $\{\mathbf e_i,\mathbf h_i\}$. Once the modes are available, the inner products \eqref{innerprod} are found, using \eqref{innerprod_psi_phi}-\eqref{fourier}. The inner products can be conveniently computed using the \textit{fast Fourier transform}.

In \eqref{fresnel} we are free to multiply each side by any factor dependent on $m=m(\text{pol},k_x,k_y)$. Ideally this will not affect the solution. However, in practice the result after pseudoinversion will differ somewhat depending on this normalization, as the different equations in the system are weighted differently. 

Given a solution $\left[r_{ij}\right]$ it is interesting to calculate the error, measured as the deviation from perfect match of the tangential fields. We assume incoming mode $i$, and denote the deviation of the transversal electric and magnetic fields by $\mathbf \Delta e_i^{(t)}$ and $\mathbf \Delta h_i^{(t)}$, respectively. We can define corresponding transformed quantities as follows:
\begin{align}
\Delta E_{im}=\frac{\kappa(m)^*}{|\kappa(m)|}\frac{1}{2}\int\limits_{\text{cell}}\mathbf \Delta e_i^{(t)}\times\mathbf{\mathcal{H}}_{m}^{(t)*}(x,y)\cdot\mathbf{\hat z}\mathrm{d}A, \\
\left(\Delta H_{im}\right)^*=\frac{\kappa(m)^*}{|\kappa(m)|}\frac{1}{2}\int\limits_{\text{cell}}\mathbf{\mathcal{E}}_{m}^{(t)}(x,y)\times\Delta\mathbf h_i^{(t)*}\cdot\mathbf{\hat z}\mathrm{d}A.
\end{align}
Denote the transmission coefficients given by \eqref{transmission_e} and \eqref{transmission_h} by $t^{(e)}_{im}$ and $t^{(h)}_{im}$ respectively. Taking $[t^{(e)}_{im}]$ as the definition of the transmission matrix, the electric field is perfectly matched. The error in the tangential magnetic field is then given by the deviation between the left and right hand side of \eqref{transmission_h}:
\begin{align}
&|\Delta H_{im}|=\left|t^{(e)}_{im}-t^{(h)}_{im}\right|.
\end{align}
Similarly we find the error of the electric field when $[t^{(h)}_{im}]$ is taken as the definition of the transmission matrix:
\begin{align}
&| \Delta E_{im}|=\left|t^{(e)}_{im}-t^{(h)}_{im}\right|. 
\end{align}
It is therefore natural to define the error
\begin{align} |\Delta t_{im}|=|t^{(e)}_{im}-t^{(h)}_{im}|.
\label{deviation_t}
\end{align} 

In this work we have chosen to work with a homogeneous ambient medium, due to the simplification that the inner product can be written as Fourier transforms when the modes $\{\mathcal{E}_m,\mathcal{H}_m\}$ are complex exponentials. Eqs.~\eqref{transmission}-\eqref{matreq} are however generally valid, and can also be applied when the transmission medium is inhomogeneous, as long as one is able to find the modes of the medium. Furthermore, Eqs.~\eqref{transmission}-\eqref{matreq} are directly applicable if one chooses to use metallic boundary to discretize the field on the waveguide side, i.e., let a metallic box of size $(2L)^2$ surround the waveguide. 

If the dimension of the waveguide is significantly smaller than the material wavelength of the light, the waveguide will be highly diffractive. In this case, the waveguide facet can be viewed as a point source, spreading spherical wavefronts. The spherical wavefronts that propagate along the plane $z=0$, are reflected at the artificial boundary, and may thus couple back into the waveguide. To prevent such artificial reflections from disturbing the solution, we may introduce some loss, i.e., the permittivity everywhere is transformed according to $\varepsilon\to\varepsilon+i\gamma\varepsilon_0$. (Here $\varepsilon_0$ is the vacuum permittivity.) The loss parameter $\gamma$ should be small enough not to alter the reflection properties of the boundary significantly. An indication of this is obtained by comparing the Fresnel reflection coefficients for reflection at an interface with index contrast $n_{\text{co}}/n_{\text{a}}$, to those of an interface with contrast $\sqrt{n_{\text{co}}^2+i\gamma}/\sqrt{n_{\text{a}}^2+i\gamma}$. The Fresnel coefficients should remain approximately unaltered for all angles of incidence. 

The modal fields and eigenvalues, $\beta^i$, of the waveguide are given from the boundary conditions and the Helmholtz equation \eqref{helmholtz} applied to e.g., the longitudinal field component $e_i^{(z)}$.
\begin{equation}
\label{helmholtz}
\left(\frac{\mathrm{d}^2}{\mathrm{d}x^2}+\frac{\mathrm{d}^2}{\mathrm{d}y^2}+n(x,y)^2-\left(\beta^i\right)^2\right)e_i^{(z)}=0
\end{equation}
Here $n(x,y)=n_{\textrm{co}}$ in the core region and $n(x,y)=n_{\textrm{cl}}$ in the cladding region. Note that by introducing the same loss in the core as in the cladding region $\left(n(x,y)^2\rightarrow n(x,y)^2+i\gamma\right)$, the Helmholtz equation \eqref{helmholtz} remains unaltered if we perform the transformation $\left(\beta^i\right)^2\rightarrow\left(\beta^i\right)^2+i\gamma$.

\section{Planar waveguide}
We will here consider the special case of a planar waveguide. This special case is interesting because the simple geometry enables us to find both radiation and bound modes analytically, and reduction from 2d to 1d offers better probabilities to test convergence criteria. It also has interesting applications, being the geometry of conventional semiconductor laser diodes.

\begin{figure}[bt]
	\centering
		\includegraphics[width=8.5cm]{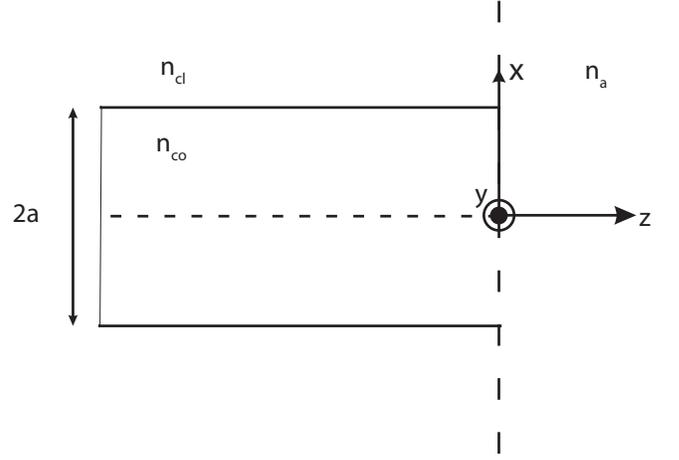}
	\caption[Geometry,planar waveguide]{Planar waveguide with thickness $2a$ oriented along $z$-axis with end facet at $z=0$. The waveguide core and cladding has refractive indices $n_{\text{co}}$ and $n_{\text{cl}}$, respectively. The ambient medium for $z>0$ is assumed to be homogeneous with refractive index $n_{\text{a}}$.}
	\label{fig:planar_wg}
\end{figure}

A planar, dielectric waveguide with thickness $2a$ is located along the $z$-axis, see Fig.~\ref{fig:planar_wg}. The refractive index of the core ($|x|<a$) and cladding ($|x|>a$) of the waveguide is $n_{\text{co}}$ and $n_{\text{cl}}$. The waveguide terminates at $z=0$; for $z>0$ there is a homogeneous nonmagnetic medium with refractive index $n_{\text{a}}$. 
We divide the modal set into two sets depending on the polarization (TE and TM). In this waveguide any TE (TM) polarized mode can be described using a superposition of TE (TM) plane waves \eqref{constvectors}. According to \eqref{orthogonal} this yields $\Psi_{im}=\Phi_{im}=0$ when $i$ and $m$ describe orthogonal polarization. Thus the two sets of modes will not couple to each other through the boundary $z=0$; so we may consider them separately. Here we will only consider TE modes.

To discretize the modes on each side of the boundary, we impose periodic boundary conditions: The fields are supposed to be periodic along the $x$-axis; for example, the fields at $x=+L$ correspond to the fields at $x=-L$. The artificial period $2L$ satisfies $2L\gg a$.

The electric field of the modes for $z>0$ can be written
\begin{equation}\label{modemetallic}
\mathbf{\mathcal{E}}_{\text{TE}}^p= A^p\sqrt{\omega\mu}\exp(ik_x^p x+ik_z^p z)\hat{\bf y}.
\end{equation}
Here $k_x^p=p\pi/L$, $\left(k_z^p\right)^2=\left(n_{\text{a}}\omega/c\right)^2-\left(k_x^p\right)^2$, and $p$ is an integer labeling the mode. The transversal component of the magnetic field is
\begin{equation}
\mathbf{\mathcal{H}}_\text{TE}^p= -\frac{A^pk_z^p}{\sqrt{\omega\mu}}\exp(ik_x^p x+ik_z^p z) \hat{\bf x}.
 \end{equation}
The normalization constants $A^p=1/\sqrt{|k_z^p|L}$ are chosen such that
\begin{equation}
\frac{1}{2}\int_{-L}^L \mathbf{\mathcal{E}}_{\text{TE}}^p\times\mathbf{\mathcal{H}}_{\text{TE}}^{p'*}\cdot\mathbf{\hat z} \mathrm{d}x=\frac{k_z^{p*}}{|k_z^p|}\delta_{pp'}.
\label{orthonormal}
\end{equation}

For the waveguide ($z<0$), the modes are a little bit more complicated. Due to reflection symmetry about the plane $x=0$, the set of modes can everywhere be divided into even modes and odd modes. These two sets will not couple to each other, so we can consider them separately. We take the $z$-dependence to be $\exp(i\beta z)$, and solve the wave equation in each region $0<x<a$ and $a<x<L$. By requiring an even mode, the electric field in the region $0<x<a$ is $B\cos(k_x' x)$, where $k_x'^2=\omega^2n_{\text{co}}^2/c^2-\beta^2$, and $B$ is a real normalization constant. For $a<x<L$ the electric field can be written $C\sin(k_xx+\varphi)$, where $k_x^2=n_{\text{cl}}^2\omega^2/c^2-\beta^2$, and $C$ and $\varphi$ are real constants. The electromagnetic boundary conditions require the electric field and its derivative with respect to $x$ to be continuous at $x=a$ and $x=L$. Moreover, the periodic boundary condition combined with even symmetry requires that $\diff \sin(k_xx+\varphi)/\diff x|_{x=L}=0$. Taken together, these conditions require that
\begin{equation}
\label{disprelTE}
k_x'\tan(k_x'a)=-k_x\tan[k_x(L-a)].
\end{equation}
Denoting the $j$-th solution of the dispersion relation \eqref{disprelTE} by superscript $j$, we can write the modal fields as follows:
\begin{equation}
 \mathbf{e}_j=\frac{\sqrt{\omega\mu}}{\sqrt{|\beta^j|}}\psi_j(x)\exp(i\beta^j z)\hat{\bf y},
\end{equation}
where
\begin{equation}\label{fieldpsi}
\psi_j(x)= 
\begin{cases} 
  D^j\cos(k_x'^j x), & \text{for } 0\le x \le a,\\
  E^j\cos[k_x^j(L-x)], & \text{for } a<x\le L,                
\end{cases}
\end{equation}
and
\begin{equation}
 E^j=D^j\frac{\cos(k_x'^j a)}{\cos[k_x^j (L-a)]}.
\end{equation}
The transversal component of the magnetic field is given by
\begin{equation}
 \mathbf{h}_j\cdot\hat{\bf x}=-\frac{\beta^j}{\sqrt{\omega\mu|\beta^j|}} \psi_j(x).
\end{equation}
The normalization constants $D^j$ are chosen to obtain ortho\-normality, similarly to \eqref{orthonormal}: 
\begin{align}
&\frac{1}{2}\int_{-L}^L \mathbf{e}_j\times\mathbf{h}_j^{'*}\cdot\mathbf{\hat z} \mathrm{d}x\\
=&\frac{1}{2}\frac{\beta^{j*}}{|\beta^{j}|}\int_{-L}^L \psi_j^{*}(x)\psi_j^{'}(x) \mathrm{d}x=\frac{\beta^{j*}}{|\beta^j|}\delta_{jj'}.
\label{orthonormalwaveguide}
\end{align}
The matrices $[\Psi_{ip}]$ and $[\Phi_{ip}]$ can be calculated using \eqref{innerprod}:
\begin{subequations}
\begin{align}\label{innerprod_planar}
					&\Psi_{ip}=\frac{A^p |k_z^p|}{\sqrt{|\beta^i|}}\frac{1}{2}\int_{-L}^{L}{\psi_i}\exp(-ik_x^p x)\mathrm{d}x\\
&\Phi_{ip}=\frac{\beta^i}{k_z^p}\Psi_{ip}.
\end{align}
\end{subequations}
From \eqref{fresnel}, the matrix Fresnel equation for a planar waveguide is
\begin{equation}\label{fresnelplanar}
\sum_j r_{ij}(k_z^p+\beta^{j})\Psi_{jp} = (\beta^{i}-k_z^p)\Psi_{ip}. 
\end{equation}
Note that for plane waves, Eq.~\eqref{fresnelplanar} is reduced to the standard Fresnel reflection coefficient for TE polarization, $r_{\text{TE}}=(k_z^{1}-k_z^{2})/(k_z^{1}+k_z^{2})$, where $k_z^i$ is the $z$-component of the propagation constant in medium $i$. 

The transmission coefficients are found from \eqref{transmission}:
\begin{subequations}\label{trans_planar}
\begin{align}
t_{ip}&=\left(\Psi_{ip}+\sum_j r_{ij}\Psi_{jp}\right), \label{trans_planar_e}\\
t_{ip}&=\frac{1}{k_z^p}\left(\beta^{i}\Psi_{ip}-\sum_j r_{ij}\beta^{j}\Psi_{jp}\right). \label{trans_planar_h}
\end{align}
\end{subequations}

Similarly to the general case in Section II, we may introduce artificial loss by modifying the refractive indices according to $n_{\text{a}}^2\to n_{\text{a}}^2+i\gamma$, $n_{\text{co}}^2\to n_{\text{co}}^2+i\gamma$, and $n_{\text{cl}}^2\to n_{\text{cl}}^2+i\gamma$. For the homogeneous medium, this implies a modification $(k_z^p)^2\to(k_z^p)^2+i\gamma\left(\omega/c\right)^2$, and for the waveguide $(\beta^j)^2\to(\beta^j)^2+i\gamma\left(\omega/c\right)^2$. The transverse wavenumbers thus remain unaltered.

We now consider a planar waveguide of width $2a$ and refractive index $n_{\text{co}}=\sqrt{10}$, completely surrounded by vacuum. The motivation for this choice of $n_\textrm{co}$ is twofold. Firstly, \textit{Vahidpour et al.} \cite{vahidpour} have previously performed calculations for this structure, giving us the possibility to confirm our numerical results. Secondly, the core refractive index is quite similar to the refractive index of GaAs, which will be considered in the next section. 

The necessary loss parameter, $\gamma$, can be reduced (increased) by increasing (decreasing) the half-width $L$ accordingly. Note however that increasing $L$ quickly becomes numerically challenging for a 2d structure, as the density of modes scales with $1/L$ in 1d and $1/L^2$ in 2d. For comparison, we have used two sets of simulation parameters; both $L=60\cdot{a}$, $\gamma=0.1$ and $L=240\cdot{a}$, $\gamma=0.025$. The latter set of parameters describes the real structure more accurate, whereas the first set of parameters have significant advantages in terms of computational resources. Comparing the Fresnel reflection coefficient at normal incidence for an interface with refractive index contrast ${n_\textrm{co}}/{n_\textrm{a}}$ with and without loss, the difference is $|\Delta r|/|r|\approx0.032$ for $\gamma=0.1$, and $|\Delta r|/|r|\approx0.008$ for $\gamma=0.025$. The waveguide was described using all modes with real $\beta$, in addition to the evanescent modes with ${\re{\beta^2}>-(10\omega{a}/c)^2}$. For half-width $L=60$, this corresponds to 577 modes at $a\omega/c$=3, and less for lower frequencies. With this high number of modes, we obtain a very good field match at the boundary. However, accurate results for the reflection matrix are obtained with fewer modes. For example, increasing the cut-off to $-(5\omega{a}/c)^2$, i.e., reducing the number of modes by approximately a factor of 2, the reflection matrix is roughly the same, while the field match at the boundary is worse.

\begin{figure}[ht]
	\centering
		\includegraphics[width=8.5cm]{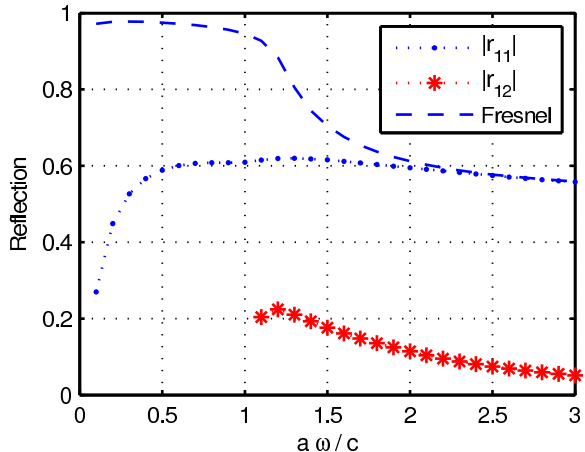}
	\caption{The reflection coefficient as a function of ${{a}\omega/c}$ for a planar waveguide of width $2a$ and refractive index $n_{\text{co}}=\sqrt{10}$. The figure shows the fraction of the fundamental mode amplitude reflected back into itself, $|r_{11}|$, and the fraction of the fundamental mode amplitude reflected into the first excited mode, $|r_{12}|$. Two sets of parameters have been used, one with $\gamma=\gamma_H=0.1$, and one with $\gamma=\gamma_L=0.025$.}
	\label{fig:rii sfa k0r}
\end{figure}

Fig.~\ref{fig:rii sfa k0r} shows the reflection of the fundamental TE mode as a function of ${\omega{a}/c}$. The figure also shows $r_{12}$, i.e. the fraction of the fundamental mode reflected into the first excited mode. The result corresponds well with results from \textit{Vahidpour et al.} \cite{vahidpour}. In the geometric optics limit, the modes can be described as plane waves with incident angles $\phi^j=\arccos(\beta^j/(n_{\text{co}}\omega/c))$. The reflection can thus be described using the Fresnel equations. This limit is indicated in Fig.~\ref{fig:rii sfa k0r}, i.e., the Fresnel reflection coefficient of a plane wave with index contrast $n_{\text{co}}/n_{a}$, and angle $\phi^j$.

Without the use of the artificial absorption, we observed oscillations in the reflection as a function of frequency. We attribute these oscillations to artificial reflection from the hard-wall boundary conditions. Different frequencies will give rise to varying waveguide-mirror resonances. By adding the artificial absorption, these oscillations were removed. The good agreement with \textit{Vahidpour et al.} \cite{vahidpour} verifies that the absorption has not altered the reflection properties significantly, even for $\gamma=0.1$.  The agreement between the calculations with $\gamma=0.1$ and $\gamma=0.025$ is also very good. In the following, we will look closer into the results from the calculation using $\gamma=0.1$. 

\begin{figure}[ht]
	\centering
		\includegraphics{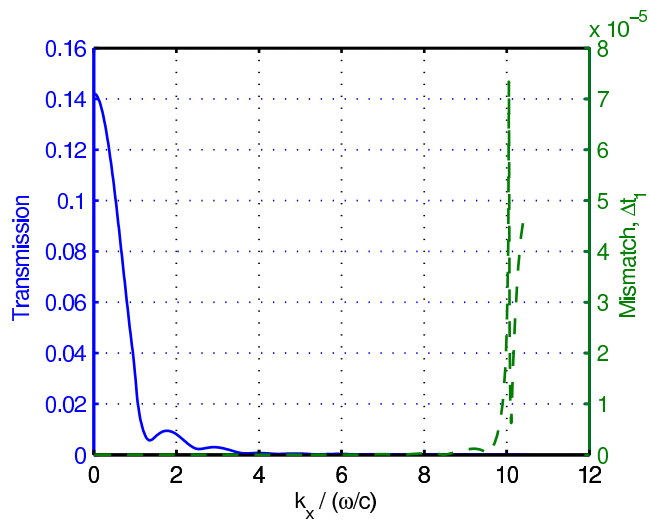}
	\caption{The transmission coefficient from the fundamental mode into plane waves with wavevectors $k_x$. The mismatch \eqref{deviation_t} is also shown.}
	\label{fig:delta_t}
\end{figure}

\begin{figure}[htb]
	\centering
		\includegraphics[width=8.5cm]{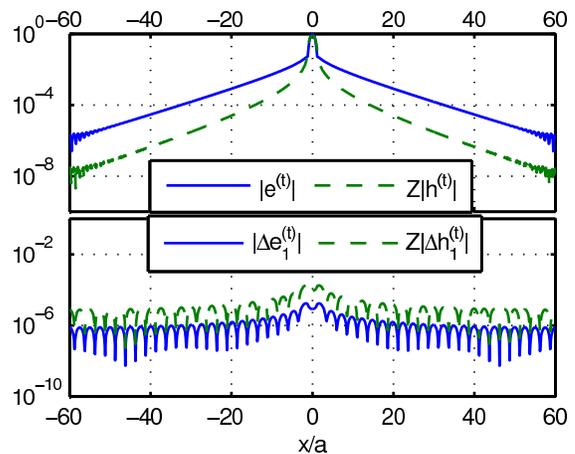}
		\caption{The transverse fields $e^{(t)}$ and $h^{(t)}$ at the boundary $z=0$, the mismatch between the fields at $z=0^+$ and $z=0^-$ is also indicated. $Z=\sqrt{\mu_0/\varepsilon_{\text{a}}}$ is the impedance of the ambient medium.}
	\label{fig:match}
\end{figure}

Fig.~\ref{fig:delta_t} shows the transmission coefficient of the fundamental mode at $\omega{a}/c$=3, calculated as the mean of the transmission matrices obtained from the two equations \eqref{trans_planar_e} and \eqref{trans_planar_h}. The deviation \eqref{deviation_t} between the transmission resulting from the two equations is also shown. The deviation increases at the spatial frequency corresponding to the highest order waveguide mode included in the model.
Fig.~\ref{fig:match} shows the match of the electric and magnetic transverse fields $e^{(t)}$ and $h^{(t)}$, with the fundamental mode incident ${a}\omega/c$=3. The match is very good, especially for the dominating spatial frequencies, i.e. the $k_x$ values where the transmission is the largest. The match can be improved further by increasing the number of modes in the calculation, i.e. by decreasing the cut-off value of $\re\beta^2$.

\begin{figure}[ht]
	\centering
		\includegraphics{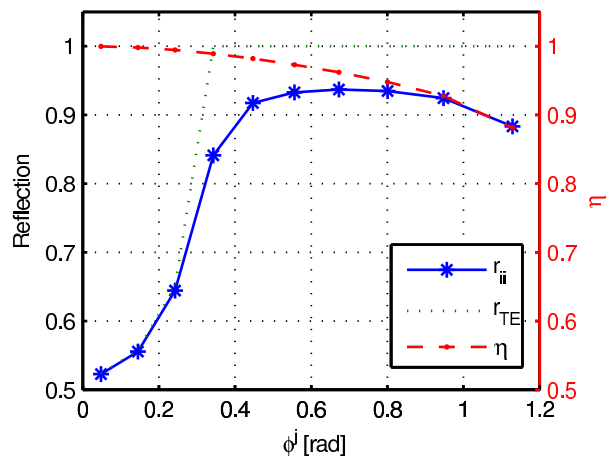}
	\caption{The reflection coefficient as a function of the modal angle $\phi^j=\arccos(\beta^j/(n_{co}\omega/c))$. For comparison the Fresnel reflection coefficient and the confinement factor $\eta$ are also shown.}
	\label{fig:rii sfa modeNo}
\end{figure}

It is interesting to compare the modal reflection coefficients with the Fresnel coefficients. From Fig.~\ref{fig:rii sfa k0r}, we see that the fundamental mode was well described by the Fresnel equations for $a\omega/c>2$. To investigate how the behavior is for higher order modes, we calculate the reflection matrix at $a\omega/c=10$. The same modal cut-off as previously was used, the loss was $\gamma=0.1$, and the half-width of the unit cell was reduced to $L=10a$. The increased normalized frequency, $a\omega/c$, leads to reduced diffraction effects, thus enabling a smaller $L$ parameter. Fig.~\ref{fig:rii sfa modeNo} shows the diagonal elements of the reflection matrix, as a function of the modal angle $\phi^j$.  

Two main factors affect the reflection in the diffraction regime. Firstly, the field is only partly confined within the waveguide, and the field outside the waveguide region experience a reduced refractive index contrast of $n_{\text{cl}}/n_{\text{a}}$ rather than $n_{\text{co}}/n_{\text{a}}$. This effect will be more pronounced with decreasing confinement, i.e., for higher order modes. The confinement factor $\eta$, defined as the fraction of power within the core, is included in Fig.~\ref{fig:rii sfa modeNo} for reference. Secondly, each mode can be described as a set of plane waves, centered around the angle $\phi^j$. Decreasing $a\omega/c$ yields an increasing angular spread of each mode. The effect of this angular spread is a smoothing of the reflection spectrum. In Fig.~\ref{fig:rii sfa modeNo}, it is seen that around the critical angle for total internal reflection the reflection is smeared out compared to the Fresnel reflection. Furthermore, for decreasing confinement factor the reflection is seen to decrease compared to the Fresnel reflection.

\section{Hexagonal nanowires}
We will now focus our attention towards two kinds of semiconductor nanowires, narrow band gap nanowires based on ZnO or GaN, and wide band gap nanowires based on GaAs. These structures have been reported to have a hexagonal cross-section. Investigations have previously been done to compare the reflection properties of hexagonal and circular nanowires \cite{henneghien}; they were reported to be approximately the same when comparing structures with the same cross-section area. It is thus convenient to describe hexagonal nanowires using an effective radius, $\rho_{\text{eff}}$ \cite{henneghien}, defined so that a circle with radius $\rho_{\text{eff}}$ has the same area as a hexagon of sides $a$, i.e., $a\approx 1.1 \times \rho_{\text{eff}}$. In our calculations we will use the exact hexagonal geometry; $\rho_{\text{eff}}$ is however chosen to describe the nanowire. This facilitates comparisons with previous work \cite{maslov, henneghien} and circular nanowires. 

The modes of the hexagonal nanowires were found using the commercial software, Comsol Multiphysics$^\text{TM}$. For these hexagonal structures it was more convenient to use a metallic rather than a periodic boundary condition to discretize the waveguide modes.  

The extension to a 2d cross-section as well as the strong diffraction present in the nanowires, impose some additional challenges compared to the planar case. To maintain a feasible model in terms of computer power and time consumption, some care was taken to limit the number of modes needed to describe the field accurately. Firstly, evanescent modes were ignored. Secondly, as the wavelength decreases the diffraction becomes less pronounced and the radial extension of the field will decrease. The half-width of the unit cell, $L$,  was thus set to be frequency dependent, decreasing linearly as a function of wavelength according to $L\omega/c=\text{constant}$.  
\subsection{Wide band gap nanowires}
We consider a hexagonal nanowire with refractive index $n_{\text{co}}=2.45$,  surrounded by vacuum. This value is close to refractive index of ZnO (2.33) \cite{yoshikawa_zno} and GaN (2.60) \cite{wang_gan} at room temperature. Such a nanowire has been studied previously by \textit{Maslov et al.} \cite{maslov} and \textit{Henneghien et al.} \cite{henneghien}. ZnO and GaN have wide bandgaps, and will typically lase around 380 nm \cite{johnson_zno} and 375 nm \cite{johnson_gan}, respectively,  at room temperature. Note that ZnO and GaN are slightly anisotropic, but this has not been taken into account.

The simulation parameter for the artificial absorption was set to $\gamma=0.1$, and the half-width of the imposed unit cell was $L=20{\rho_\text{eff}}$ at the lowest frequency, and decreasing as a function of increasing frequency.

\begin{figure}[ht]
	\centering
	\includegraphics{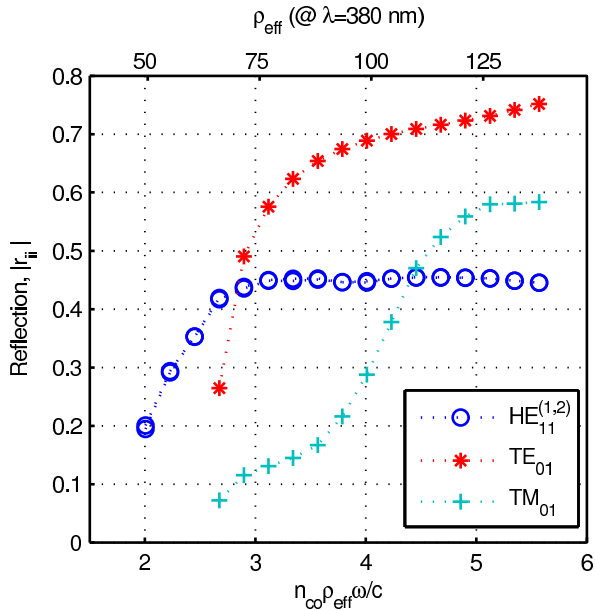}
	\caption{The reflection coefficient for the lowest order modes for a hexagonal nanowire with $n_\textrm{co}=2.45$ surrounded by vacuum, $n_{\text{cl}}=1$. The superscript designate the degenerate fundamental modes HE$_{11}^{(1)}$ and HE$_{11}^{(2)}$.}
	\label{fig:r_i_i_enkel_hex_zno}
\end{figure}

Fig.~\ref{fig:r_i_i_enkel_hex_zno} shows the diagonal elements of the reflection matrix for a nanowire with $n_{\text{co}}=2.45$, for the 3 modes with the lowest cut-off. The next excited mode has its cut-off value at $\rho_\text{eff}\omega/c=1.5$; a higher number of modes should be included in the calculation in order to describe the reflection properties of such higher order modes accurately. We used around 220 waveguide modes to describe the field in the nanowire. The result coincide quite well with results reported earlier \cite{maslov,henneghien}, especially for the fundamental mode ($\text{HE}_{11}$). The reflection coefficient deviates slightly more for the TE$_{01}$ and TM$_{01}$ mode, especially for the TM$_{01}$ in the frequency regime where this mode has a very low confinement. 

After a regime where the reflection is dominated by the reduced modal confinement, each mode approach the Fresnel reflection as observed in the planar case. The fundamental mode is in this region close the Fresnel reflection at normal incidence for a refractive index contrast $2.45/1$, i.e 0.42. For the higher order modes $\text{TE}_{01}$ and $\text{TM}_{01}$ the average incident angle of the plane waves constituting the modes will be higher, so that in the geometric optics picture one would have total internal reflection. The reflection is however limited by the reduced confinement. This effect explains the fact that the reflection of these modes rise quickly, to exceed the reflection of the fundamental mode. Highest possible reflection coefficient and single mode operation are important to optimize a laser. This is achieved for a radius around 60 nm for the nanowire considered here, i.e., just below cut-off of the first excited mode. In this region the modal reflection coefficient of the fundamental mode is 0.35. For larger radii, the TE$_{01}$ mode will start to dominate the modal spectrum due to its much higher reflection coefficient, whereas for smaller radii the reflection coefficient will decrease due to limited confinement.

\begin{figure}[ht]
	\centering
		\includegraphics[width=8.5 cm]{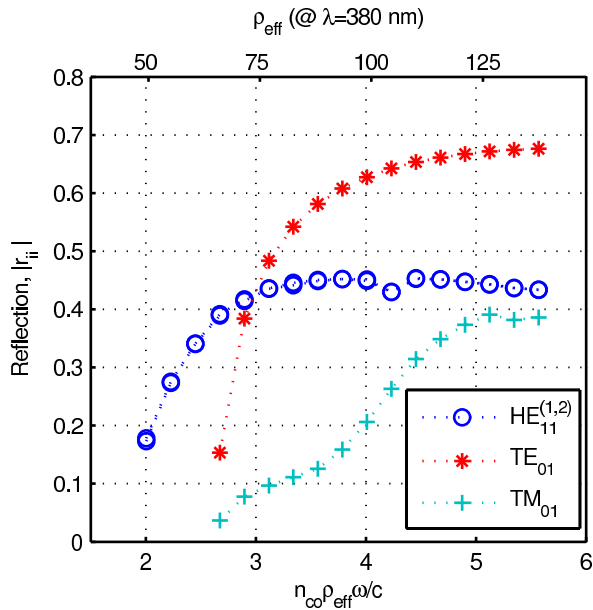}
			\caption{Rough estimate of the reflection coefficient for the lowest order modes for a ZnO hexagonal nanowire surrounded by vacuum, using only bound modes.}
	\label{fig:r_i_i_enkel_hex_zno_kun_bundne}
\end{figure}

If one is interested only in the qualitative behavior of the fundamental mode, it is possible to obtain a quick and quite good estimate of the reflection coefficient using only bound modes. The result of such an estimate is shown in Fig.~\ref{fig:r_i_i_enkel_hex_zno_kun_bundne}.
The resemblance to the more accurate calculations (Fig.~\ref{fig:r_i_i_enkel_hex_zno}) is quite remarkable for the fundamental mode. Note that all that is needed is to Fourier transform less than 10 modes. One might simplify further by approximating the nanowire as a cylindric waveguide; in this case the modes could be described using analytic expressions from standard textbooks \cite{snyder}.

\begin{figure}[hbt]
	\centering
		\includegraphics[width=8.5cm]{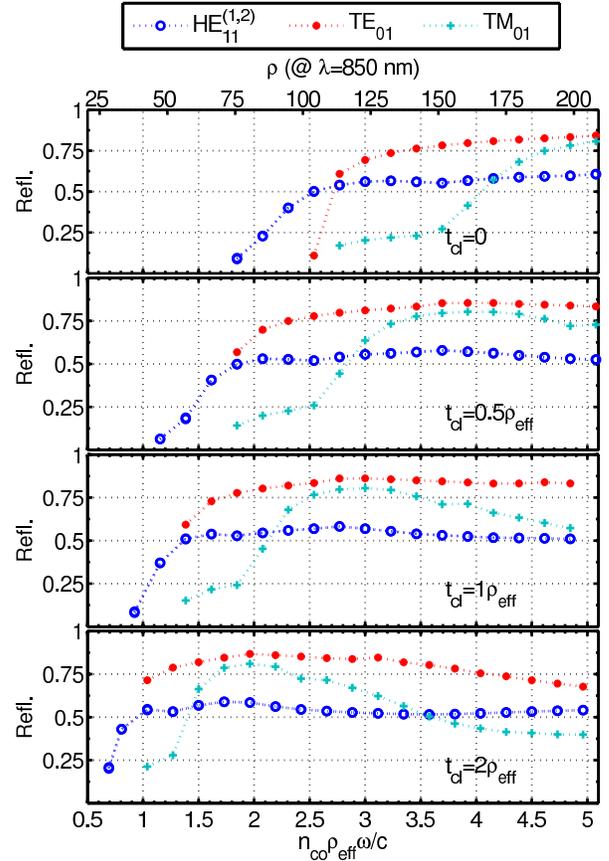}
	\caption[cladding series]{Reflection coefficient of the three lowest order modes for a series of GaAs based nanowires, with core refractive index $n_{\text{co}}=3.3$. The cladding consists of Al$_{0.3}$Ga$_{0.7}$As shells, $n_\text{cl}=3.15$, of varying thickness $t_{\text{cl}}$. The waveguides are surrounded by vacuum.}
	\label{fig:gaAs_series}
\end{figure}

\subsection{GaAs-based hexagonal nanowires}
GaAs has a more narrow band-gap compared to ZnO, and GaAs will typically lase around 850-870 nm at room temperature. As will be seen, nanowire lasers will thus need to have a larger diameter compared to the wider band gap nanowire considered in the previous section. Surface defects are often a problem for nanowires due to the high surface to volume ratio of nanowires. To reduce the amount of surface defects, it might be advantageous to grow a radial shell surrounding the nanowire. Such a shell will serve as a cladding material of the nanowire waveguide. A radial shell of AlGaAs \cite{zhou:AlGaAs} or GaAsP \cite{Takashi:GaAsP} is commonly used. We have investigated how the reflection properties vary with increasing cladding thickness for an Al$_{0.3}$Ga$_{0.7}$As shell. Fig.~\ref{fig:gaAs_series} shows the reflection for one pure GaAs nanowire, and three GaAs/Al$_{0.3}$Ga$_{0.7}$As core-shell nanowires, all surrounded by vacuum. We have used refractive index values for infrared wavelengths, i.e. the core refractive index is $n_{\text{co}}=3.3$, while the refractive index of the cladding is $n_{\text{cl}}=3.15$. 

The artificial loss parameter was again taken to be $\gamma=0.1$. For the pure GaAs nanowire, the factor $\rho_{\text{eff}}/L$ was 20 for the lowest frequency, linearly decreasing for increasing frequency. The nanowires with cladding had the same cell sizes as the pure wire. To better enable comparison between the two nanowire systems, we have used a scaled abscissa, $n_\text{co}\rho_{\text{eff}}\omega/c$, taking into account the difference in refractive index. Comparing Fig.~\ref{fig:r_i_i_enkel_hex_zno} and the pure nanowire in Fig.~\ref{fig:gaAs_series}, we see that the qualitative behavior is very similar, but that the higher refractive index of GaAs leads to an increased reflection. In the geometric optics limit, the fundamental mode of the pure GaAs nanowire should approach the Fresnel reflection at normal incidence, which is 0.53. From Fig.~\ref{fig:gaAs_series}, we see that the fundamental mode reflection is close to this value from $n_{\text{co}}\omega/c\rho_{\text{eff}}=2.5$. To optimize the lasing performance, one should operate the laser at a wavelength where the nanowire waveguide is single mode, and with the highest possible fundamental mode reflection. It is also desirable to achieve lasing at the smallest possible diameter. Operating the nanowire laser at 850 nm, the effective radius of a pure GaAs nanowire should be around 100 nm. As could be expected, the increase in reflection compared to ZnO or GaN based nanowires is proportional to the Fresnel reflection at normal incidence, 0.53/0.42, and the optimal radius has increased with a factor equal to the ratio of material wavelengths at the lasing frequency.  As the cladding thickness increases we see a clear trend in terms of reflection properties. Higher reflection can be obtained for smaller radii, but the reflection curve of the fundamental mode also has a higher gradient in the single mode region. It thus becomes more difficult to design an efficient nanowire laser, due to shrinking of the single mode region where the reflection of the fundamental mode is high. Another feature that is important when choosing the cladding thickness is the fraction of the modal field confined in the active GaAs core \cite{hauschild,maslov:modalGain}, to maximize gain.

We now consider the pure GaAs nanowire in the region around the single mode cut-off. If some of the design parameters deviate slightly, one might excite several modes. It is therefore useful to know how the modes couple to each other in this region, and how excitation of higher order modes can be diagnosed by analyzing the far field.

\begin{figure}[ht]
	\centering
		\includegraphics{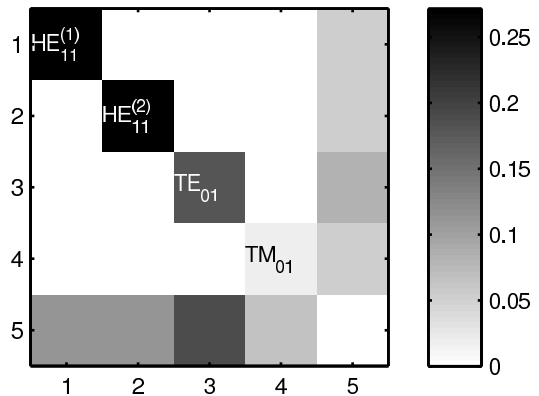}
			\caption{The power reflection matrix $\left[|r_{ij}|^2\right]$ for the bound modes of a GaAs hexagonal nanowire, with $\rho_{\text{eff}}=60$~nm and  $n_{\text{co}}\omega/c\rho_{\text{eff}}=2.5$. The last element of each row/column displays the total reflected energy from/to the given bound mode to/from radiation modes.}
	\label{fig:r_mat_GaAs_pure}
\end{figure}

Fig.~\ref{fig:r_mat_GaAs_pure} shows the reflection matrix at $n_{\text{co}}\omega/c\rho_{\text{eff}}=2.5$, i.e., just above the single mode cut-off. The reflection matrix figure shows the power reflection for the bound modes, $\left[|r_{ij}|^2\right]$; the last element of each row/column displays the total reflected energy from/to the given bound mode to/from radiation modes. From the reflection matrix we see that the coupling between the various bound modes is quite weak, whereas there is a significant coupling to radiation modes. Cross coupling between bound modes would have strengthened the competing excited modes at the expense of the fundamental modes, and thus e.g. deteriorate the beam profile.

\subsection{Far field}
The beam profile in the far field is an important aspect of the lasing performance. The far field radiation of a circular ZnO nanowire laser has previously been investigated numerically by \textit{Maslov et al.} \cite{maslov:farField}. An advantage of our method is that the far field is directly obtainable from the transmission matrix, so no further simulations are necessary. 

\begin{figure}[htb]
	\centering
		\includegraphics[width=8.5cm]{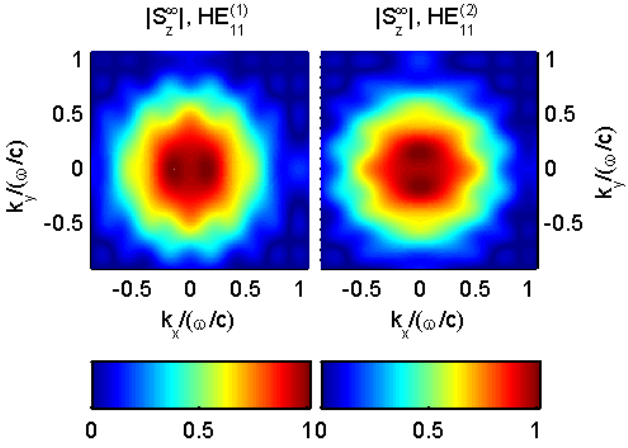}
	\caption[Far field,HE]{Far field, $|S_z^\infty|$, for the pure GaAs nanowire excited by the degenerate pair of fundamental modes; $n_{\text{co}}\rho_\text{eff}\omega/c=2.75$.}
	\label{fig:far_field_Sz}
\end{figure}

\begin{figure}[htb]
	\centering
		\includegraphics[width=8.5cm]{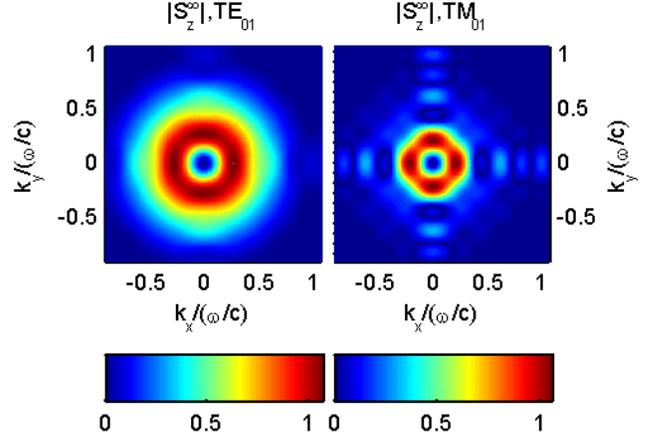}
	\caption[Far field,TE]{Far fields, $|S_z^\infty|$, for the pure GaAs nanowire excited by the modes $\text{TE}_{01}$ and $\text{TM}_{01}$; $n_{\text{co}}\rho_\text{eff}\omega/c=2.75$.}
	\label{fig:far_field_Sz_exc}
\end{figure}

For the calculations of the reflection and transmission, as mentioned previously, we utilized an artificial absorption to avoid reflections from the imposed boundary conditions. The reflection and transmission coefficients of the actual structure were thus approximated by the reflection and transmission coefficients of the structure with loss $\gamma$. Once $[r_{ij}]$ and $[t_{im}]$ have been found, one should again consider the actual structure. The artificial absorption should thus not be included in the far field calculations, i.e., the medium for $z>0$ is isotropic and homogeneous with real refractive index $n_{\text{a}}$. The far field is given from the Fourier transform \cite{hecht} of the field at $z=0$, which can be conveniently found from the transmission matrix $[t_{im}]$. For incoming mode $i$, the Fourier transform of the electric field at the interface is simply $\tilde{\mathbf{E}}_i(k_x,k_y)=\hat{\mathbf{\mathcal{E}}}_{m(k_x,k_y,\text{TE})}t_{i,m(k_x,k_y,\text{TE})}+\hat{\mathbf{\mathcal{E}}}_{m(k_x,k_y,\text{TM})}t_{i,m(k_x,k_y,\text{TM})}$, and similarly for the magnetic field. As a measure of the beam profile we calculate the $z$-component of Poynting's vector. In the far field, at the point specified by direction $[k_x,k_y,k_z]$, this power flux is
\begin{align}
|S_z^{\infty}(k_x,k_y)|&\propto k_z^2\left(|t_{i,m(k_x,k_y,\text{TE})}|^2+|t_{i,m(k_x,k_y,\text{TM})}|^2\right).
\end{align}
We investigated the far field of the pure GaAs nanowire at $n_{\text{co}}\rho_{\text{eff}}\omega/c=2.75$, using the simulation data from Fig.~\ref{fig:gaAs_series}.
Due to periodic boundary conditions, the far field becomes very coarse. We therefore performed an interpolation by padding with zeros in the Fourier domain to restore sufficient spatial resolution.
Fig.~\ref{fig:far_field_Sz} shows $|S_z^{\infty}(k_x,k_y)|$, when the fundamental mode is incident to the end facet of the pure GaAs nanowire. Any presence of higher order incoming modes, in addition to the fundamental mode, will lead to a change in the far field. For reference, the far fields for incoming modes $\text{TE}_{01}$ and $\text{TM}_{01}$ are therefore shown in Fig.~\ref{fig:far_field_Sz_exc}. The qualitative behavior is in accordance with previously reported data \cite{maslov:farField}. Note that upon reflection of i.e. the fundamental mode, higher order modes will also be excited and give contributions to the far field. Fig.~\ref{fig:far_field_Sz} is thus not simply the far field of the fundamental mode, but of the total sum of modes excited upon reflection of the fundamental mode. Also note that the hexagonal cross-section is visible in Fig.~\ref{fig:far_field_Sz}. If the fundamental mode is excited in the nanowire, the far field beam profile will be Gaussian like, with a peak intensity in the forward direction ($k_x=k_y=0$). The two higher order modes will on the other hand lead to zero emission in the forward direction. 

\section{Multimode optical fiber}
\begin{figure}[ht]
	\centering
		\includegraphics[width=8.5cm]{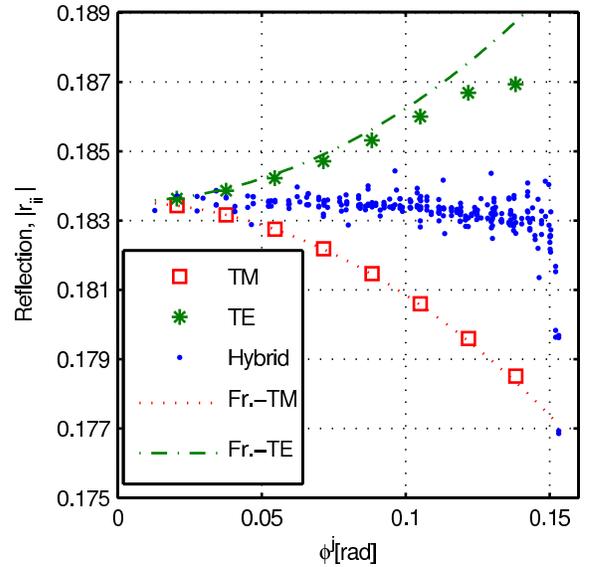}
			\caption{The diagonal reflection matrix elements, $|r_{ii}|$, for the bound modes of a silicon step-index optical fiber of radius $\rho=12.5~\mu$m, terminated in air.}
	\label{fig:r_ii_opt_fiber_a}
\end{figure}

\begin{figure}[ht]
	\centering
		\includegraphics[width=8.5cm]{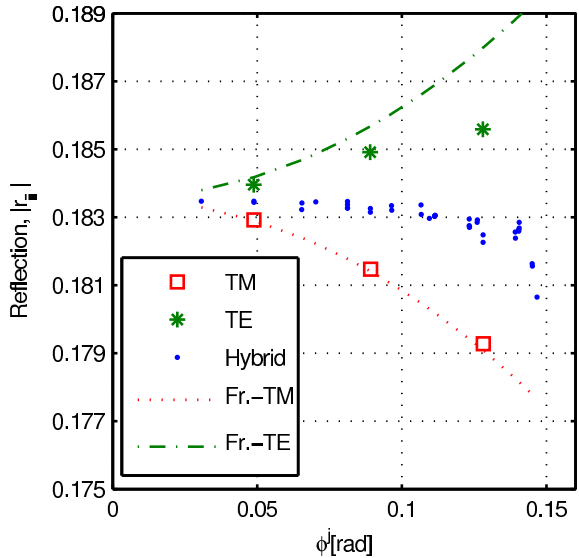}
			\caption{The diagonal reflection matrix elements, $|r_{ii}|$, for the bound modes of a silicon step-index optical fiber of radius $\rho=5~\mu$m, terminated in air.}
	\label{fig:r_ii_opt_fiber_b}
\end{figure}

Reflection properties of multimode optical fibers are interesting in order to understand the modal composition of the light after reflection. We have calculated the facet reflection of two commercially available silicon step index optical fibers from Thorlabs, with $n_{\text{co}}=1.4570$ and $n_{\text{cl}}=1.4537$ at the wavelength 633 nm. The core radii of the two fibers are $\rho=12.5~\mu$m and $\rho=5~\mu$m. The cladding thickness of both fibers is 125$~\mu$m, being so thick that we have approximated the claddings to be infinite. These multimode fibers have $\omega\rho/c\approx 125$ and $\omega\rho/c\approx 50$, so the diffraction effects are less pronounced than those of the structures we have studied previously. Therefore the unit cell can be reduced compared to the nanowires. We have here used $L=10\rho$, and no artificial absorption was introduced. Furthermore, we have neglected coupling to radiation modes; this may cause a reduced accuracy for the highest excited modes. The modes of the optical fiber were found using the analytical expressions for cylindrical waveguides \cite{snyder}. 
The diagonal elements of the reflection matrix for light at the wavelength 633 nm are shown in Figs.~ \ref{fig:r_ii_opt_fiber_a}-\ref{fig:r_ii_opt_fiber_b}. The reflection is plotted as a function of the modal angle $\phi^j=\arccos(\beta^j/({n_{\text{co}}\omega/c}))$.

The Fresnel reflection coefficients $r_{\text{TE}}$ and $r_{\text{TM}}$ for plane waves with incident angle $\phi^j$, are also shown in Figs.~\ref{fig:r_ii_opt_fiber_a}-\ref{fig:r_ii_opt_fiber_b}. Note that in analogy to the observations for the planar waveguide, the reflection coefficient of the modes with TE, TM polarization resemble the $r_{\text{TE}}$, $r_{\text{TM}}$ coefficients, respectively. The polarization properties of the hybrid modes can be seen as a mixture of TE and TM polarization, and they have reflection coefficients intermediate between the $r_{\text{TE}}$ and $r_{\text{TM}}$ coefficients. The comparison to the Fresnel reflection is best for the lowest order modes (with smallest angles $\phi^j$). Similarly to the planar case this can be explained by the reduced confinement of the higher order modes. Note however that the accuracy is somewhat decreased for the highest order modes, due to the neglect of radiation modes. 

For these optical fibers, the intermodal coupling is very weak. The maximum coupling from the fundamental mode into the other modes, $\max\left(|r_{1,j}|/|r_{1,1}|\right)$, for these two optical 
fibers was $1.3\times 10^{-3}$ and $4.1\times 10^{-4}$ for the fiber with $\rho=5 ~\mu$m and $\rho=12.5 ~\mu$m, respectively. As expected, there is more modal cross coupling in the smaller fiber, where diffraction effects are more pronounced.

\section{Conclusion}
We have presented a general formalism for calculating the end facet reflection in various waveguide geometries. The method is a natural generalization of the Fresnel equations to corresponding matrix equations, describing the reflection and transmission into all modes. A special focus in this article has been on highly diffractive waveguides, and more specifically nanowire waveguides, where the end facet reflection is vital for the nanowire lasing performance. The end facet reflection was studied for ZnO nanowires, and for GaAs/Al$_{0.3}$Ga$_{0.7}$As nanowire heterostructures with various cladding thicknesses. In general the reflection of the fundamental mode is significantly different from the plane wave Fresnel reflection for normal incidence. However, for both systems the maximum reflection in the single mode region was found to be close to this value. As a by-product, the model provide the field at the boundary, so that the far field beam profile can be directly calculated. The model is also well suited for calculations on multimode fibers, as we have demonstrated for two commercial fibers. For all waveguide geometries, we demonstrate that the reflection from a waveguide end facet is mainly governed by two factors, the modal field confinement and the Fresnel reflection of the plane waves constituting the waveguide mode.
\begin{acknowledgments}
This work was supported by the "NANOMAT" program (Grant No. 182091) of the Research Council of Norway. 
\end{acknowledgments} 
%
\end{document}